# Observation of in-gap surface states in the Kondo insulator SmB$_6$ by photoemission

J. Jiang*[1], S. Li*[2], T. Zhang*[1], Z. Sun[3], F. Chen[3], Z. R. Ye[1], M. Xu[1], Q. Q. Ge[1], S. Y. Tan[1], X. H. Niu[1], M. Xia[1], B.P. Xie[1], Y. F. Li[2], X. H. Chen[3], H. H. Wen[§2], D. L. Feng[∥1]

[1] State Key Laboratory of Surface Physics, Department of Physics, and Advanced Materials Laboratory, Fudan University, Shanghai 200433, China
[2] National Laboratory of Solid State Microstructures and Department of Physics, Nanjing University, Nanjing 210093, China
[3] Department of Physics, University of Science and Technology of China, Hefei, Anhui, 230026, People's Republic of China
* These authors contributed equally to this work.
∥email: dlfeng@fudan.edu.cn   §email: hhwen@nju.edu.cn

Kondo insulators (KI's) are strongly correlated materials in which the interactions between *4f* and conduction electrons lead to a hybridization gap opening at low temperature [1-2]. SmB$_6$ is a typical KI, but its resistivity does not diverge at low temperatures, which was attributed to some "in-gap" states [3-10]. However after several decades of research, the nature and origin of the in-gap states remain unclear. Recent band calculation and transport measurements suggest that the in-gap states could actually be ascribed to topological surface states. SmB$_6$ thus might be the first realization of topological Kondo insulator (TKI) [13], the strongly correlated version of topological insulator (TI) [11,12]. Here by performing angle-resolved photoemission spectroscopy (ARPES), we directly observed several dispersive states within the hybridization gap of SmB$_6$, which cross the Fermi level and show negligible $k_z$ dependence, indicative of their surface origin. Furthermore, the circular dichroism (CD) ARPES results of the in-gap states suggest the chirality of orbital momentum, and temperature dependent measurements have shown that the in-gap states vanish simultaneously with the hybridization gap around 150 K. These strongly suggest their possible topological origin.

The electronic structure of $SmB_6$ is characterized by narrow *4f* bands and *5d* conduction bands. At high temperature, the many-body interactions between local spins and conduction electrons, *i.e.* the Kondo screening, are weak, and the system behaves like a metal. With decreasing temperature, an energy gap develops due to the hybridization of *4f* bands and conduction *5d* bands [3], featured by a rapidly rising resistance. But there is a long-standing puzzle: its resistivity does not diverge but saturates at temperatures below 4K (ref. 4-6, see Fig. 1a). The residual conductivity was attributed to some "in-gap" states, evidenced by optical, neutron scattering, specific heat, and ARPES experiments [7-9, 18]. Although various models have been proposed to address the in-gap states [5,10], their exact nature remains illusive after decades of studies.

Recently, the novel concept of TKI may shed new light on the in-gap states in $SmB_6$ [13,14]. In analogy to topological insulators defined in weakly interacting systems [11,12], Kondo insulators with strong correlations can also be topologically classified. A topologically nontrivial KI could be realized by strong spin-orbit coupling associated with a hybridization band gap [13,14]. If $SmB_6$ is a TKI as predicted [13-16], it will possess gapless surface states that are protected by time-reversal symmetry, which naturally explains the origin of "in-gap" states and the residual conductivity. Several recent experiments have observed pronounced surface-dominating transport in $SmB_6$ [18-22], suggesting the TKI scenario. However, direct electronic structure evidences for the topological surface states have not been reported.

In this letter we present an ARPES study of $SmB_6$ single crystal. Sample and experimental details are in the Methods section and supplementary materials. The resistivity of our $SmB_6$ sample exhibits a sharp increase at T < ~50 K and starts to saturate at T < 5 K (Fig. 1a), similar to previous reports [4-6]. Fig. 1b shows the low-energy electron diffraction (LEED) image of the cleaved surface. Clean 1×1 pattern of the (001) surface is observed. The bulk and projected Brillouin zone of the (001) surface are shown in Fig. 1c. Figs. 1d and 1e present the photoemission intensities measured along $\overline{\Gamma}$-$\overline{X}$ and $\overline{X}$-$\overline{M}$ directions over a large energy scale. Three flat bands could be observed at 18meV, 150meV, and 950 meV below the

Fermi energy ($E_F$), as marked in the integrated energy distribution curves (EDC's). They are the $^6H_{5/2}$, $^6H_{7/2}$ and $^6F$ multiplets of the Sm$^{2+}$ $4f^6 \to 4f^5$ final state, respectively [18,23]. In addition, there is a broad but highly dispersive band centered at $\overline{X}$ (dashed line). In previous ARPES studies [23] it was ascribed to the bulk $5d$ band based on the band calculations [17].

To search for the in-gap states in SmB$_6$, the photoemission intensity at $E_F$ is collected over the surface Brillouin zone (Fig. 2a). Data were taken at 8 K, well into the KI phase. Indeed, four large oval-shaped Fermi surfaces are observed around $\overline{X}$, together with some spectral weight at $\overline{\Gamma}$, and the intensity around $\overline{\Gamma}$ is much enhanced in the second Brillouin zone. The nature of these states is further illustrated in a cut across $\overline{X}$ in Fig. 2c. Here the hybridization gap is manifested by the suppressed spectral weight between $E_F$ and the $^6H_{5/2}$ $4f$ band at -18meV (referred as ϕ). But there is an "in-gap" dispersive band centered at $\overline{X}$ (referred as α), which gives the large Fermi surface. The dispersion of α is clearly visible in the momentum distribution curves (MDC's) in Fig. 2d as well, with Fermi crossings at ±0.29 Å$^{-1}$. The band becomes obscure when it crosses ϕ and disappears at about -30meV. On a larger energy scale, the aforementioned $5d$ like band is marked as δ in Fig. 2b, and the band between the two flat $4f$ bands is referred as β. The photoemission intensity along a cut across $\overline{\Gamma}$ is shown in Fig. 2e, where the vague feature located at $\overline{\Gamma}$ exhibits electron-like band dispersion as further illustrated by their MDC's in Fig. 2f. The electron-like dispersion is more clearly seen in the second Brillouin zone as shown in Fig. 2g and Fig. 2h. This band (referred as γ) gives the tiny central electron pocket in Fig. 2a. The bright spot on ϕ at $\overline{\Gamma}$ reflects the crossing of the two γ branches at 23 ± 3 meV. Furthermore, an oval shaped Fermi surface α' can be observed around $\overline{\Gamma}$ which shows almost the same size as the α Fermi surface around $\overline{X}$, and is more visible in the second Brillouin zone. It seems to be a shadow Fermi surface from the umklapp scattering of the α Fermi surface around $\overline{X}$ by the reported 1×2 surface reconstruction[18], however, the superstructure spot of such a reconstruction is missing in the LEED pattern of our sample. Probably it is too weak for the sensitivity of our LEED apparatus. Another possible cause could be the poor $k_z$ resolution of the photoemission spectroscopy as will be discussed later, if the α band is of bulk origin. Since there is an X point on top of Γ along the $k_z$ direction for a cubic crystal, the residual weight of the α pocket will be observed around $\bar{\Gamma}$. However, this is less likely since the bulk SmB$_6$ is an insulator.

As predicted in recent first principle calculations [15,16], SmB$_6$ is a possible TKI with surface states at $\bar{X}$ and $\bar{\Gamma}$ points. Since the bulk SmB$_6$ is an insulator based on the transport data and calculations, the in-gap states α and γ should be some metallic surface states. To further illustrate this, we performed photon-energy dependent measurements to reveal the $k_z$ dependence of the electronic structure. The photoemission intensity distribution across $\bar{X}$ and the MDC at $E_F$ taken with different photon energies are presented in Figs. 3a and Fig. 3b respectively. The exploited photon energies cover a full $k_z$ period in the extended Brillouin zone (Fig. 3c), and the dispersions and Fermi crossings of α do not show any noticeable $k_z$ dependence. Moreover, Fig. 3d shows the photon energy dependence of large energy scale photoemission intensity distribution across $\bar{X}$. Since $k_x$, $k_y$, and $k_z$ are equivalent in a cubic compound, the bulk *5d* bands are expected to have similar fast dispersions along $k_z$. However, at the first sight, the *5d*-like fast dispersive bands together with others do not show strong $k_z$ dependence either, and the envelope of the photoemission intensity map in Fig. 3d seems to be independent of photon energy. However, the bright interior of the 29 eV and 31 eV data are typical consequence of the projection of intensities from different $k_z$'s due to the poor $k_z$ resolution of ARPES [24]. As is further illustrated by the broad MDCs of the *5d* band at $\bar{X}$ in Fig. 3e, the finite $k_z$ resolution of ARPES smears out the intensity of distribution for a band structure with strong $k_z$ dispersions. The MDC's for the α band in Fig. 3b are much sharper than those in Fig. 3e, so it is originated from the surface, while the β and δ bands are most likely originated from the bulk 5d band.

To study whether these surface states are topologically nontrivial or not, the chirality of their spin and orbital momentum shall be examined, since it is a remarkable hallmark of the topological surface state. Spin-resolved ARPES requires long acquisition time due to its low count rate, and at this stage, it is not feasible for the weak surface state signal discussed. Alternatively, instead of examining the chirality of the spin, one could examine the chirality of the orbital momentum with the so-called circular dichroism (CD) ARPES experiment. This technique has been demonstrated to be a powerful tool to investigate the topological origin of the surface state by measuring the circular dichroism of the surface state[25,26]. Figs. 4a and 4b show the Fermi surface mapping with right circularly polarized (RCP) light and left circularly polarized (LCP) light respectively. Comparing the two data sets, we can clearly see the difference between the two Fermi surfaces taken with two different circular polarizations. The intensity is higher in the upper electron pocket on the positive $k_y$ side than in the lower electron pocket on the negative $k_y$ side for RCP, while for the LCP, it shows a totally opposite behavior. The electron pocket at the right side also shows opposite intensities between the positive $k_y$ side and negative $k_y$ side with different circularly polarized light. However, the electron pocket at the left side does not show much CD, which is probably due to some matrix element effects and requires further theoretical simulation. The CD is more clearly demonstrate by the difference between RCP and LCP data in Fig. 4c, where the LCP data was subtracted from the RCP data. The intensity of the α Fermi surface is antisymmetric with respect to the horizontal axis, which resembles the CD observed in the topological insulator $Bi_2Se_3$ before[25]. As for the γ band at $\overline{\Gamma}$, it is very hard to figure out any differences in Fig. 4a-4c since the intensity of γ band is quite weak. However, there apparently exists certain intensity inversion between the RCP and LCP data in the MDCs across $\overline{\Gamma}$ along the $k_y$ direction in Figs. 4d and 4e. We can see higher intensity on the negative $k_y$ side for the RCP data but on the positive $k_y$ side for the LCP data. Therefore, both the α and γ band show similar CD that resembles the nontrivial surface state of a topological insulator, which indicates that both the α and γ band might be topologically nontrivial. The differential mapping of RCP and LCP data in Fig. 4c shows that for each α pocket the lower part is positive and the upper part negative. Similar behavior also holds for the γ pocket, as suggested by Figs. 4d and 4e. This finding indicates that each pocket of surface states in $SmB_6$ bear the same chiral property in momentum space.

If α and γ are topological surface states of a topological Kondo insulator, they should vanish when the hybridization gap closes at high temperatures in a KI. Figs. 5a present temperature dependent ARPES intensity measured across $\overline{X}$. We found that although these bands are surface ones, they still exhibit the crossover from a TKI to a metal with increasing temperature. α stays almost unchanged at T<80K, but its band velocity starts to increase above 80 K. Eventually it appears that α and β merge into one highly dispersive band at 150 K, which fits the large-scale band dispersion shown in Fig. 1e. Therefore, α, β, and δ could be the broken sections of the *5d* band induced by the *4f-5d* hybridization. However, one also sees that α goes straight through ϕ (Figs. 2b-c), which suggests another possibility that α might contain or coexist with a topological surface state, although it disappears before ending at a Dirac point. The temperature evolution of the hybridization can be further demonstrated by the integrated EDC's of the images in Fig. 5a. After some processing described in the caption of Fig.5, the resulting EDC's are presented in Fig. 5b. As temperature increases, the peak near -20 meV (representing ϕ) is gradually suppressed, and the dip near $E_F$ (reflecting the hybridization strength) is gradually filled. This indicates that the *4f* states become localized and completely lose coherence at T ≈ 150 K. Consequently, they are decoupled with the *5d* bands and the hybridization disappears [27,28]. Similarly in the photoemission intensity near $\overline{\Gamma}$ plotted in Fig. 5c, the γ band also seems to disappear at 150 K, which can be visualized by the MDC's in Fig. 4d as well. Since there are no bulk dispersive bands around $\overline{\Gamma}$ in the calculations [15,16], the γ band is likely a topologically nontrivial surface state that vanishes with the hybridization gap. Moreover, since the (001) surface Brillouin contains one $\overline{\Gamma}$ and four $\overline{X}$ points, if γ is topologically nontrivial, no matter α is trivial or not, there will be totally odd number of surface state which indicates the whole system is topologically nontrivial.

Unlike in the case of weak correlated systems, where band calculations worked quantitatively well for describing TIs [11,12], there are only qualitative agreements between our data and the calculations of SmB$_6$, such as the existence of surface states at $\overline{\Gamma}$ and $\overline{X}$. Moreover, different calculations could differ significantly, depending on how strong the correlations were treated [15,16]. The surface state at $\bar{X}$ was predicted to give a rounded hole pocket in one calculation [16], but a very anisotropic electron pocket in another [15], while the observed α Fermi surface is an oval-shaped electron pocket with 5-6 times area of the calculated ones. The discrepancies might be partially because the actual cleaved surface is different from the ideal surface used in the calculations. Furthermore, they also demonstrate that correlation would affect the topological surface states significantly.

In summary, we present direct observation of the in-gap surface states in $SmB_6$. The circular dichroism, and photon energy and temperature dependences of these in-gap states are systematically studied, which support that $SmB_6$ is a possible TKI. Our findings lay the foundation for understanding the anomalous transport properties of $SmB_6$ and further exploration of the interplay between strong correlations and topological effects in TKI's in general.

**Methods:**

High quality single crystals of $SmB_6$ were grown with the arc melting technique by establishing a temperature gradient in the slow cooling process. At first, the metal Sm were cut into small pieces and mixed with the boron powder in the ratio of Sm:B=1:6. The mixture was pressed into a pellet with the diameter of 1 cm and thickness of about 0.5 cm. Then the pellet was heated up in the water-cooled clean oven for arc melting. Welding was repeated for five times in order to ensure the sample uniformity. The output is a bulk that contains many shiny crystals in the sizes of millimeter or sub-millimeter.

The samples were cleaved at the (001) plane and measured under ultra high vacuum of $3\times10^{-11}$ *torr*. The in-house ARPES measurements were performed with SPECS UVLS discharge lamp (21.2 eV He-Iα light). The synchrotron ARPES experiments were performed at the Beamline 5-4 of Stanford synchrotron radiation lightsource (SSRL), and the one-cubed ARPES station of BESSY II. The circular dichroism experiments were performed at beamline 9A of Hiroshima Synchrotron Radiation Center. All the data were taken with Scienta R4000 electron analyzers, the overall energy resolution was better than 7 meV, and the angular resolution was 0.3 degree.

**Acknowledgements**

The authors are grateful for the discussions with Prof. R. B. Tao and Prof. G. M. Zhang, and the experimental support by Dr. D. H. Lu, Dr. Y. Zhang and Dr. M. Hashimoto at SSRL, Dr. E. Rienks at BESSY II, and Dr. M. Arita at Hisor. This work is supported in part by the National Science Foundation of China, and National Basic Research Program of China (973 Program) under the grant Nos. 2012CB921400, 2011CB921802, 2011CBA00100, 2012CB821403, 2012CB21400. SSRL is operated by the US DOE, Office of Basic Energy Science, Divisions of Chemical Sciences and Material Sciences.

**Author contributions**

J.J., Z.R.Y., Z. S., S.Y.T., M.Xu, Q.Q.G., X.H.N., M.Xia, and B.P.X. performed ARPES measurements. S.L. and Y.F.L. grew most of the samples and conducted

sample characterization measurements, F. C. and X. H. C. grew the samples for the CD-ARPES measurement, J.J., T.Z., and D.L.F. analyzed the ARPES data, T.Z., D.L.F., J.J. and H.H.W wrote the paper. D.L.F. and H.H.W. are responsible for the infrastructure, project direction and planning.

## Additional Information

The authors declare no competing financial interests. Correspondence and requests for materials should be addressed to D.L.F. (dlfeng@fudan.edu.cn) and H.H.W. (hhwen@nju.edu.cn).

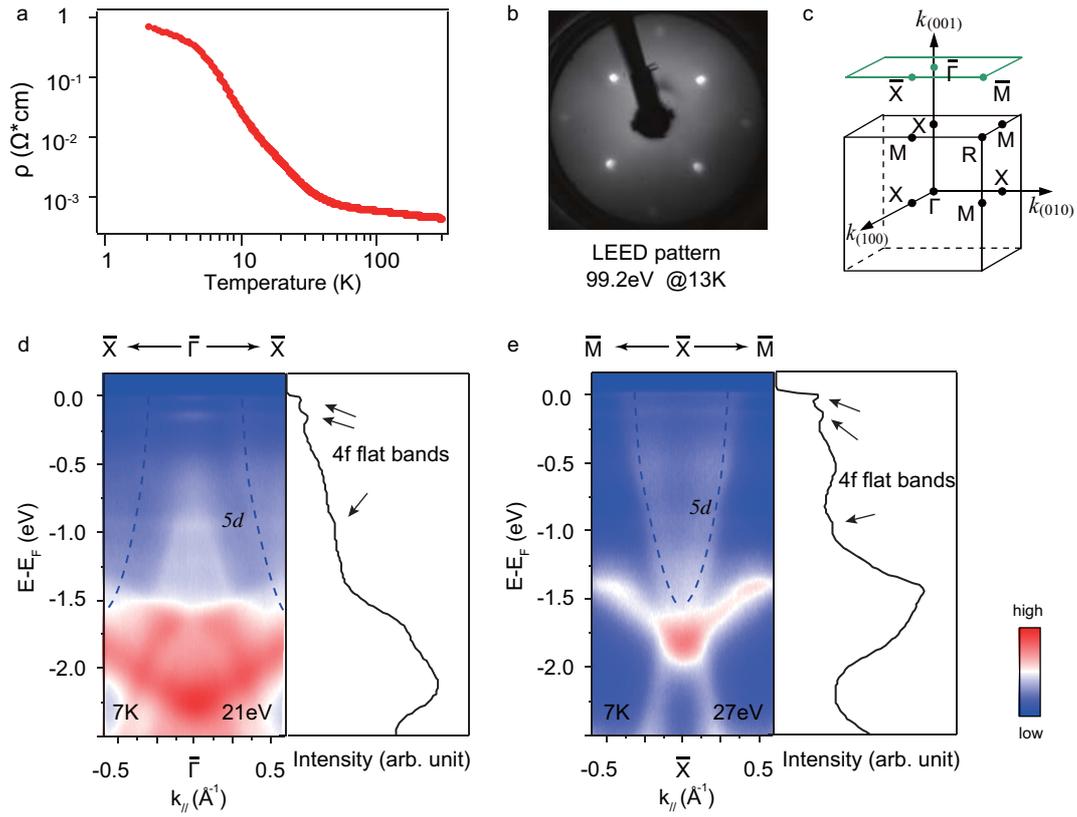

**Figure 1 | SmB$_6$ sample characterization and valence band structure. a**, Temperature dependence of the resistivity of SmB$_6$. **b**, Low-energy electron diffraction (LEED) pattern of the cleaved SmB$_6$ (001) surface. Bright spots in square lattice reflect the pristine 1×1 surface, with the lattice constant of 4.13Å. **c**, Bulk and surface projected Brillouin zone of SmB$_6$ and the high symmetry points. **d** and **e**, Left panel: Large energy scale photoemission intensity plot along **(d)** the $\bar{\Gamma}$-$\bar{X}$ and **(e)** $\bar{X}$-$\bar{M}$ direction. Right panel: integrated energy distribution curves (EDC's) to highlight the positions of the flat *4f* bands as marked by the arrows. Dashed lines in **d** and **e** indicate the Sm *5d* band that crosses $E_F$ at high temperatures. Data were taken at 7K with 21eV and 27eV photons respectively in SSRL.

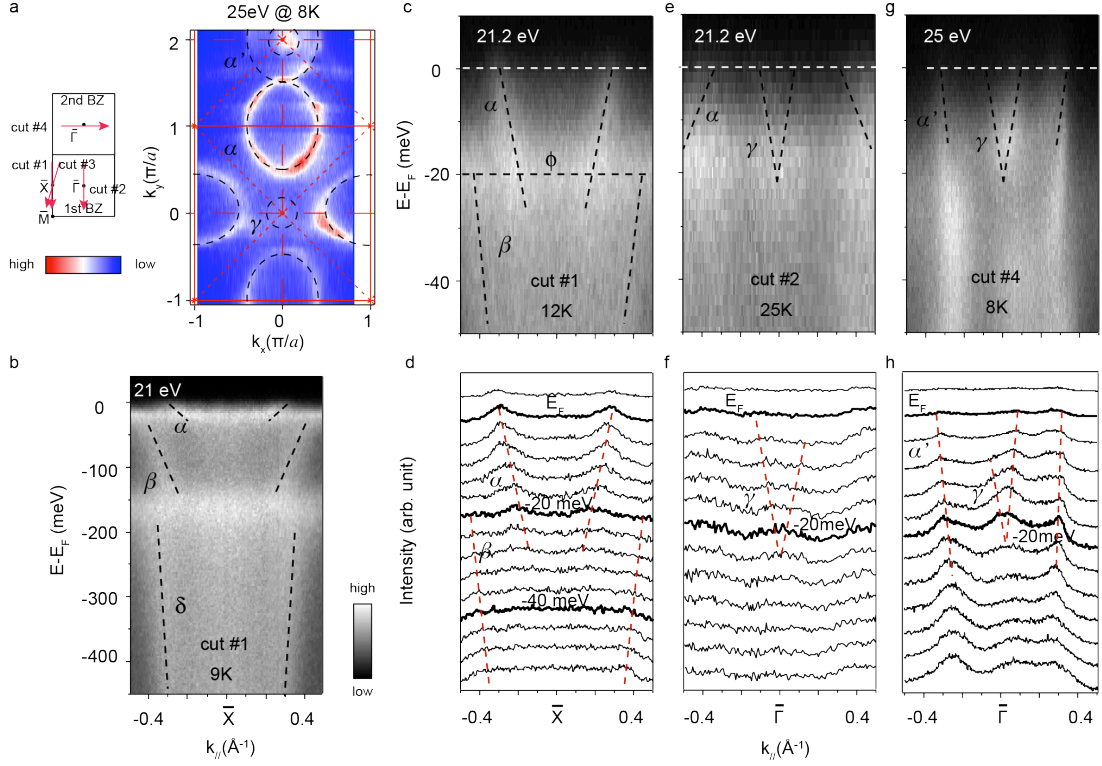

**Figure 2 | Dispersive in-gap states around $\overline{\Gamma}$ and $\overline{X}$. a**, Photoemission intensity map at the Fermi energy for SmB$_6$ taken with 25 eV photons in Hisor. The intensity was integrated over a window of (E$_F$-5meV, E$_F$+5meV). **b**, A larger energy scale photoemission intensity plot along cut #1 with 21 eV photons in SSRL. c, The photoemission intensity, and **d**, momentum distribution curves (MDC's) along cut #1 in the projected two-dimensional Brillouin zone as shown in the left of panel **a**. **e**, The intensity plot and **f**, MDC's along cut #2. All the spectrum in panel **c~f** are taken with He-Iα (21.2eV) photons. **g**, The intensity plot and **h**, MDC's along cut #4 with 25 eV photons in Hisor. The dashed lines in panels **b-h** just schematically show the dispersions of various bands, ignoring band warping at crossings due to hybridization. The temperatures that the data were taken are labeled in individual panels. The α and γ Fermi surfaces cover ~24 % and ~1 % of the surface Brillouin zone, respectively. By assuming a linear dispersion, the Fermi velocities of α and γ are 0.24 and 0.22 eV Å, respectively; their (extrapolated) crossing points are located at 65, and 23 meV below E$_F$ respectively.

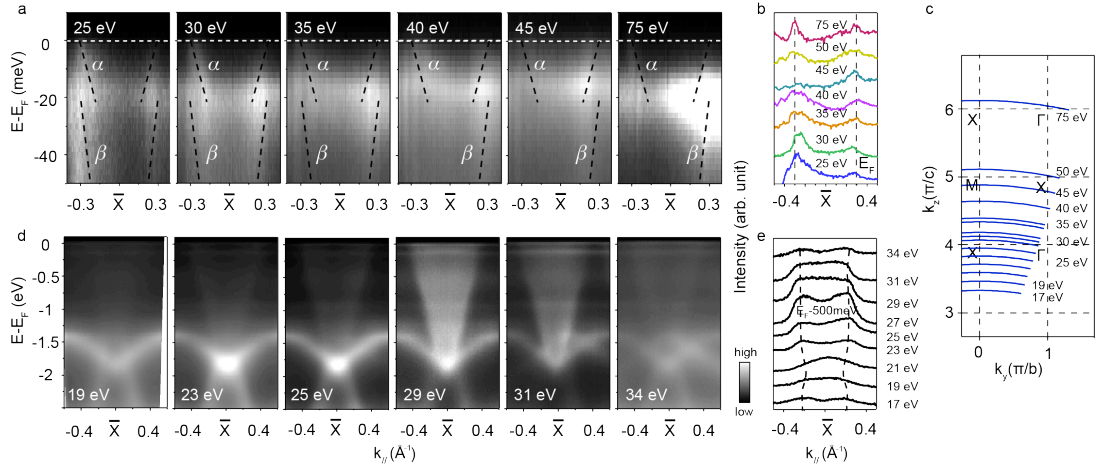

**Figure 3 | Photon energy dependence of the in-gap state around $\overline{X}$. a,** Photon energy dependence of the photoemission intensity measured along cut #3 in Fig.2a. The α and β band dispersions are tracked by the dashed lines. The bright regions correspond to where the bands hybridize. They are particularly strong at some photon energies due to matrix element effects. Data were taken at 1 K at BESSY II. **b**, MDC's at $E_F$ integrated over [$E_F$ - 6meV, $E_F$] taken from data in panel **a**. **c**, The correspondence between all the photon energies used and the sampled momentum cuts, indicating that the photon range used cover a full period of $k_z$ in the extended Brillioun zone. **d**, Photon energy dependence of the photoemission intensity measured along cut #1 in Fig.2a. Data were taken at 7 K at SSRL. **e**, MDC's at $E_F$-500meV taken from data in panel **d**.

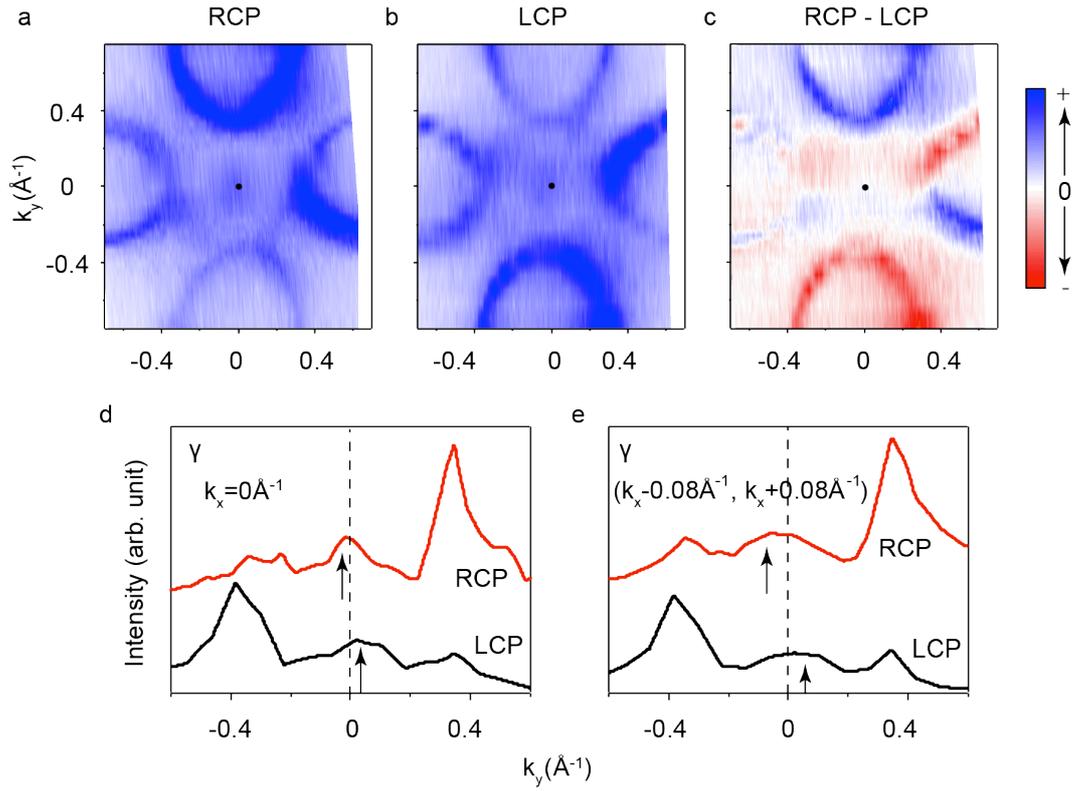

**Figure 4 | Circular dichroism in the surface states at $\overline{X}$ and $\overline{\Gamma}$. a** and **b,** Fermi surface mapping with right circularly polarized (RCP) light and left circularly polarized (LCP) light respectively. The intensity was integrated over a window of ($E_F$-5meV, $E_F$+5meV). **c**, the difference between the RCP and LCP photoemission intensity map. **d** and **e**, difference between the RCP and LCP of the γ band, the MDCs along $k_y$ are not integrated in panel **d** while they're integrated in a window of ($k_x$-0.08Å$^{-1}$, $k_x$+0.08 Å$^{-1}$) in panel **e**. All the data were taken at 8K with 25eV photons at Hisor.

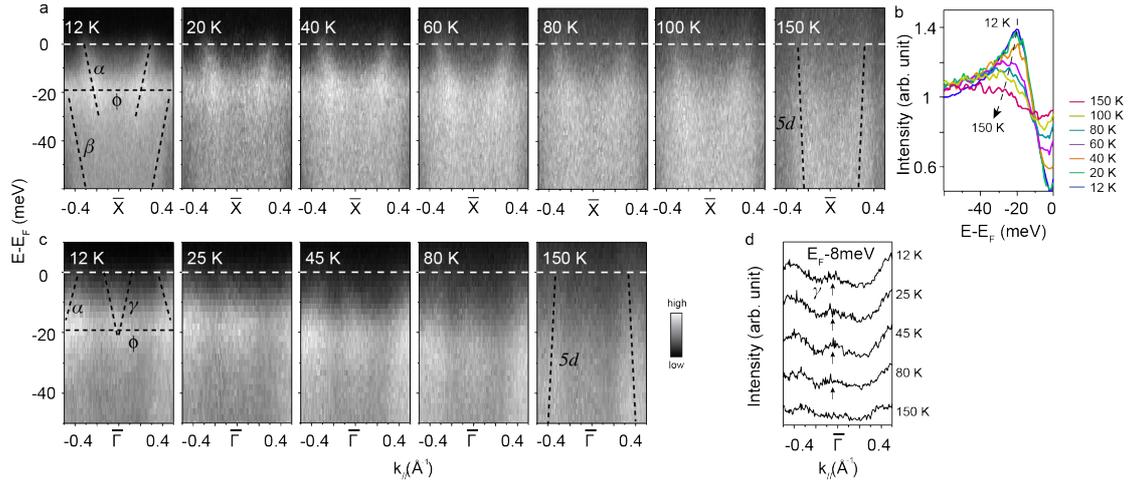

**Figure 5 | Temperature dependence of the photoemission data of the surface state around $\overline{\text{X}}$. a,** Photoemission intensity of SmB$_6$ along cut #1 in Fig. 2a at 12 K, 20 K, 40 K, 60K, 80 K, 100 K, and 150 K respectively. **b,** Temperature dependence of the angle-integrated spectrum, obtained from the data in panel **a**, after dividing the energy-resolution convoluted Fermi-Dirac distribution functions at individual temperatures to remove the temperature broadening effects and recover density of states near $E_F$. **c,** Photoemission intensity of SmB$_6$ along cut #2 in Fig. 2a at 12 K, 25 K, 45 K, 80 K, and 150 K respectively. **d,** Temperature dependence of the MDC at 8 meV below $E_F$ for data in panel **c**, integrated over $\pm 2$ meV. All data were taken with 21.2 eV photons.

Supplementary information for

Observation of in-gap surface states in the Kondo insulator SmB$_6$ by photoemission

J. Jiang*[1], S. Li*[2], T. Zhang*[1], Z. R. Ye[1], M. Xu[1], Q. Q. Ge[1], S. Y. Tan[1], X. H. Niu[1], M. Xia[1], B.P. Xie[1], Y. F. Li[2], H. H. Wen[§2], D. L. Feng[∥1]

[1] State Key Laboratory of Surface Physics, Department of Physics, and Advanced Materials Laboratory, Fudan University, Shanghai 200433, China
[2] National Laboratory of Solid State Microstructures and Department of Physics, Nanjing University, Nanjing 210093, China
* These authors contributed equally to this work.
∥email: dlfeng@fudan.edu.cn  §email: hhwen@nju.edu.cn

**Sample characterization:**

To check the structure and the crystallinity of the single crystals, we have ground the crystals into powder for measuring the X-ray diffraction (XRD). As shown in Fig. S1, the XRD looks very pure without any secondary phase. The solid line here is the Rietveld fitting using the software of TOPAS, which gives a fitting result of 100 % of the SmB$_6$ phase with negligible standard deviation.

In order to illustrate the Kondo effect in the system, we measured the magnetoresistance as shown in Fig. S2 and Fig. S3. One can see that the magnetoresistance is negative in low temperature region, while it gradually becomes positive in high temperature region (T≥20 K). The negative magnetoresistnace can be understood as the cause of the Kondo effect. The Kondo singlet will be partially broken by the magnetic field and thus release the localized d-band electrons, leading to the decreasing of resistance. In the high temperature region, the positive magnetoresistance is induced by the orbital scattering effect.

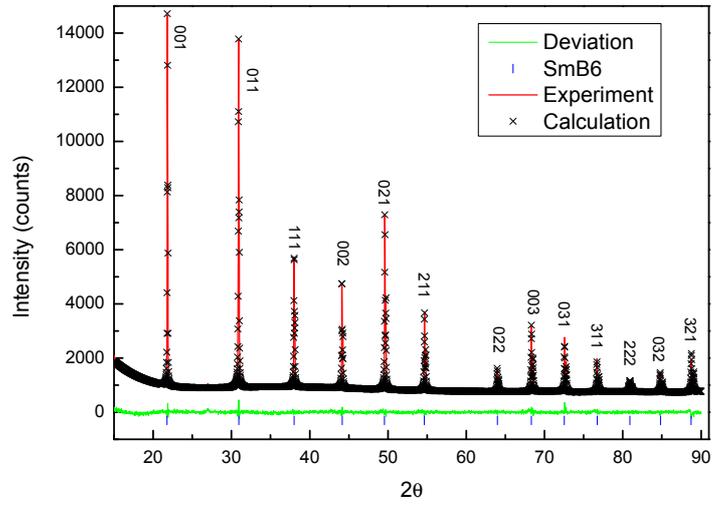

**Figure S1**: Powder X-ray diffraction pattern of $SmB_6$ crystal. Solid line is the Rietveld fitting using the software of TOPAS. The standard deviation is negligible.

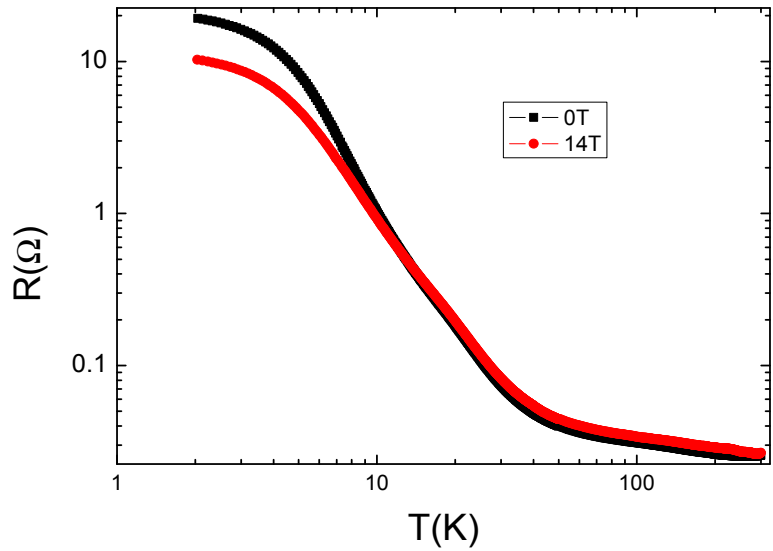

**Figure S2**: The temperature dependence of resistance at zero field and 14 T. A negative magnetoresistance is observed in the low temperature region, while it becomes positive in the high temperature region.

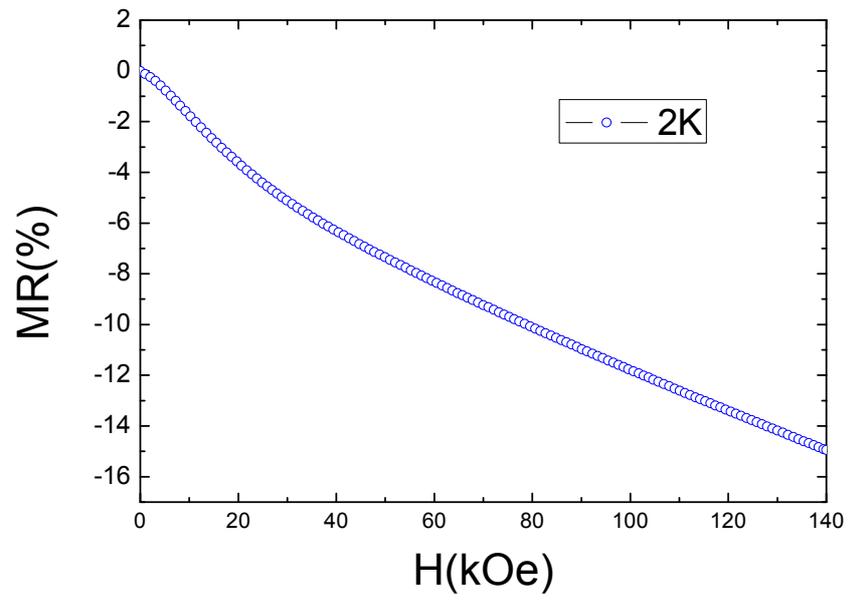

**Figure S3**: The magnetoresistance of $SmB_6$ taken at 2 Kelvin. A negative magnetoresistance is obvious here.